\documentclass[aps,pra,twocolumn,amsmath,floatfix,showpacs,superscriptaddress]{revtex4}

\usepackage{graphicx}
\usepackage{color}
\usepackage{marvosym}
\usepackage{wasysym}
\usepackage{pifont}
\usepackage{amssymb}
\usepackage{amsmath}
\usepackage{amsfonts}
\usepackage{epsfig}
\usepackage[american]{babel}
\usepackage{bm} 

\begin{document}
\newcommand{\matel}[3]{\langle #1 | #2 | #3 \rangle}
\newcommand{\ud}{\mathrm{d}}
\newcommand{\spinor}{\vec{\Psi}}
\newcommand{\wf}{\Psi}
\newcommand{\vect}[1]{{\bf{#1}}}
\newcommand{\fmatrix}{{\bm{\mathcal{F}}}}
\newcommand{\vv}[1]{\tilde{#1}}
\newcommand{\ie}{{\rm i.e.}}
\newcommand{\eg}{{\rm e.g.}}
\newcommand{\etal}{{\rm et al.}}

\title{Stabilization and pumping of giant vortices in dilute Bose--Einstein condensates}
\author{Pekko Kuopanportti}
\email{pekko.kuopanportti@tkk.fi}
\affiliation{Department of Applied Physics/COMP, Aalto University, P.O. Box 14100, FI-00076 AALTO, Finland}
\author{Mikko M\"ott\"onen}
\affiliation{Department of Applied Physics/COMP, Aalto University, P.O. Box 14100, FI-00076 AALTO, Finland}
\affiliation{Low Temperature Laboratory, Aalto University, P.O. Box 13500, FI-00076 AALTO, Finland}
\date{\today}

\begin{abstract}
Recently, it was shown that giant vortices with arbitrarily large quantum numbers can possibly be created in dilute Bose--Einstein condensates by cyclically pumping vorticity into the condensate. However, multiply quantized vortices are typically dynamically unstable in harmonically trapped nonrotated condensates, which poses a serious challenge to the vortex pump procedure. In this theoretical study, we investigate how the giant vortices can be stabilized by the application of a Gaussian potential peak along the vortex core. We find that achieving dynamical stability is feasible up to high quantum numbers. To demonstrate the efficiency of the stabilization method, we simulate the adiabatic creation of an unsplit 20-quantum vortex with the vortex pump.
\end{abstract}

\hspace{5mm}


\maketitle

\section{\label{sc:intro}Introduction}

Bose--Einstein condensation in trapped gases of alkali-metal atoms was observed experimentally in 1995~\cite{Anderson1995,Bradley1995,Bradley1997,Davis1995}. A few years later, those pioneering experiments were followed by the creation of singly quantized vortices~\cite{Matthews1999,Madison2000a} and vortex lattices~\cite{Madison2000b,Abo-Shaeer2001,Raman2001} in such systems. Since quantized vortices manifest the long-range phase coherence of the condensates and are inherently connected with superfluidity, there has been wide interest in studying their properties in trapped Bose--Einstein condensates (BECs)~\cite{Fetter2009}. In particular, the stability of vortices has been the subject of intensive research~\cite{Dodd1997,Rokhsar1997,Pu1999,Isoshima1999,Svidzinsky2000,Virtanen2001,Simula2002,Kawaguchi2004,Jackson2005,Huhtamaki2006a,Lundh2006,Capuzzi2009,Kuopanportti2010a}.

In principle, a quantized vortex in a BEC can carry any number of circulation quanta. However, it is well known that typically a vortex with a winding quantum number greater than unity has a higher energy than the corresponding number of separated single-quantum vortices, which means that multiquantum vortices have a propensity to split into single-quantum vortices~\cite{Mottonen2003,Shin2004,Huhtamaki2006b,Mateo2006,Isoshima2007,Takahashi2009,Kuopanportti2010b,Kuwamoto2010}. Although this result is generally true only in an infinite homogeneous system, it still holds in finite-sized BECs for a majority of trap geometries and particle numbers. Being able to create vortices with large winding numbers would provide access to novel vortex-splitting patterns beyond the typical linear chain that prevails in the decay of two- and four-quantum vortices~\cite{Shin2004,Isoshima2007}. Because of the distinct nature of the different splitting patterns predicted for highly quantized vortices~\cite{Kuopanportti2010b}, observing the decay of such states would enable a lucid comparison between theory and experiment. It has also been speculated that the splitting of multiquantum vortices may create necessary conditions for the initialization of superfluid turbulence~\cite{Aranson1996}.

The energetic instability renders the creation of multiquantum vortices challenging but not impossible. In fact, energetically unstable states can be quite long-lived since the relaxation to lower-energy states necessitates dissipation which comes mainly in the form of noncondensed atoms. Temperatures in typical experiments are low enough such that the noncondensed component of the gas is negligible and, consequently, relaxation is slow. Therefore, methods that do not rely on the relaxation of condensate dynamics are especially well-suited for producing multiquantum vortices. In topological phase imprinting~\cite{Nakahara2000,Isoshima2000,Ogawa2002,Mottonen2002}, the condensate remains in its instantaneous eigenstate, and the process is insensitive to the exact rate of change and path of the control fields. 

Although topological phase imprinting was the first experimentally realized means to create multiquantum vortices~\cite{Leanhardt2002}, it was only recently~\cite{Mottonen2007} when it was shown to be applicable in creating giant vortices with essentially arbitrarily large winding numbers. In this \emph{vortex pump},  a fixed amount of vorticity is added to the condensate in each adiabatic pumping cycle, and thus in principle, arbitrarily large vorticities can be reached. However, a serious challenge is posed by the \emph{dynamical instability} of the giant vortices that becomes more pronounced as the vorticity increases~\cite{Huhtamaki2006a,Kuopanportti2010a}. Dynamical instabilities can lead to dissociation of the vortex even in the absence of dissipation~\cite{Mottonen2003,Shin2004,Huhtamaki2006b,Mateo2006,Isoshima2007,Takahashi2009,Kuopanportti2010b,Kuwamoto2010}, and therefore their effect cannot be disposed of by reducing temperature. 

Motivated by the above considerations, we study how vortices with large winding numbers can be made dynamically stable in nonrotated harmonically trapped BECs. As our method of choice, we investigate the effect of applying a repulsive Gaussian-shaped plug potential along the symmetry axis of the trap. In the vortex pump, the plug not only serves to stabilize the vortex but also prevents unwanted spin flips in the central region of the trap. The plug potential can be realized in experiments by a tightly focused far-blue detuned laser beam~\cite{Abo-Shaeer2001,Raman2001,Simula2005,Davis2009,Neely2010} as was done already in the pioneering work of Davis~\etal~\cite{Davis1995}.

This article is structured as follows. In Sec.~\ref{sc:theory}, we present the mean-field theory of BECs and relate the  computational parameters of the plug potential to the properties of the laser. We also describe how the vortex pump is modeled.  In Sec.~\ref{sc:results}, we present our numerical results and show that vortices up to high winding numbers can be stabilized with experimentally achievable plug potentials. We demonstrate the efficiency of the plug by simulating the creation of an unsplit 20-quantum vortex with the vortex pump. Section~\ref{sc:discussion} is devoted to discussion.

\section{\label{sc:theory}Theoretical and numerical methods}

We consider a BEC in a nonrotated, cylindrically symmetric harmonic trap and restrict the analysis to the zero-temperature limit, thereby ignoring the effects of the thermally excited atoms~\cite{Griffin2009}. Experiments with dilute BECs can usually be carried out at temperatures where this approximation is justified~\cite{Leanhardt2003b}. In the stability analysis below, we assume that the spin degree of freedom of the BEC is fixed by the Zeeman coupling to a strong external magnetic field. Under these circumstances, the scalar order parameter field $\wf$ of the BEC is described by the time-dependent Gross--Pitaevskii (GP) equation,
\begin{equation}\label{eq:tGPE}
i\hbar\partial_t\wf(r,\phi,t) = \left[ {\mathcal H}+ g|\wf(r,\phi,t)|^2 \right] \wf(r,\phi,t),
\end{equation}
where ${\mathcal H}$ denotes the single-particle Hamiltonian and the atom-atom interaction strength $g$ is related to the vacuum $s$-wave scattering length $a$, the atomic mass $m$, and the axial harmonic oscillator length $a_z=\sqrt{\hbar/\left(m\omega_z\right)}$ by $g=\sqrt{8\pi}\hbar^2 a/\left(m a_z\right)$. In Eq.~(\ref{eq:tGPE}), we have assumed that the condensate is pancake-shaped, \ie, that the harmonic trapping frequency in the axial direction is much greater than in the radial direction, $\omega_z \gg \omega_r$, which has enabled us to factor out the $z$ dependence of the full order parameter as $\wf_\mathrm{full} (\vect{r},t)=\wf(r,\phi,t)\exp[-z^2/\left(2a_z^2\right)]/\sqrt[4]{\pi a_z^2}$. The single-particle Hamiltonian is defined by
\begin{equation}\label{eq:Ham}
{\mathcal H} = -\frac{\hbar^2}{2m}\left(\partial_r^2+\frac{1}{r}\partial_r + \frac{1}{r^2}\partial^2_\phi \right) + V(r),
\end{equation}
where the potential function includes the possible optical plug potential, $V(r)=m\omega_r^2 r^2/2 +V_\mathrm{plug}(r)$. The order parameter is normalized such that $\int |\wf|^2 r\,\ud r \ud\phi = N$, where $N$ is the number of condensed atoms. 

Stationary states of the system satisfy the time-independent GP equation, which is obtained from Eq.~(\ref{eq:tGPE}) with the replacement $i\hbar\partial_t \longrightarrow \mu$, where $\mu$ is the chemical potential. Here, the stationary states are chosen to be axisymmetric vortex states with a winding number $\kappa$, implying that the order parameter can be written in the form
\begin{equation}\label{eq:psi}
\wf(r,\phi) = \sqrt{n(r)}e^{i\kappa\phi},
\end{equation}
where $n(r)$ is the areal particle density.

Small-amplitude oscillations about the stationary states play an important role in the study of BECs~\cite{Pethick2008}. We decompose the order parameter as
\begin{equation}\label{eq:decomposition}
\wf(r,\phi,t)=e^{-i\mu t/\hbar}\left[\wf(r,\phi)+\chi(r,\phi,t)\right],
\end{equation}
where we assume the oscillatory part $\chi$ to have a small $L^2$ norm compared with $\wf$. In the Bogoliubov theory, one seeks a solution in the form
\begin{equation}\label{eq:oscillation}
\chi=\sum_q \left[ u_q(r)e^{i(\kappa+l_q)\phi-i\omega_q t} + v_q^\ast(r)e^{i(\kappa-l_q)\phi+i\omega_q^\ast t} \right],
\end{equation}
where the complex-valued functions  $u_q$ and $v_q$ are the quasiparticle amplitudes corresponding to the index $q$. Each quasiparticle mode is also characterized by an integer $l_q$ that determines the angular momentum of the mode with respect to the condensate. By substituting Eqs.~(\ref{eq:decomposition}) and (\ref{eq:oscillation}) into Eq.~(\ref{eq:tGPE}), neglecting terms superlinear in $\chi$, and noting that $\wf(r,\phi)$ satisfies the time-independent GP equation, we arrive at the Bogoliubov equations
\begin{equation}\label{eq:bogo}
\left( \begin{array}{cc} {\mathcal{L}}_{\kappa+l_q} & gn(r)  \\ -g n(r) & -{\mathcal{L}}_{\kappa-l_q}\end{array} \right) \left( \begin{array}{c} u_q(r) \\ v_q(r) \end{array} \right) = \hbar \omega_q \left( \begin{array}{c} u_q(r) \\ v_q(r) \end{array} \right),
\end{equation}
where 
\begin{equation}
{\mathcal{L}}_{\kappa}= -\frac{\hbar^2}{2m}\left(\partial_r^2+\frac{1}{r}\partial_r - \frac{\kappa^2}{r^2}\right) + V(r) -\mu + 2g n(r).
\end{equation}

The Bogoliubov excitation spectrum $\{\omega_q\}$ can be used to classify the stability of the corresponding stationary state. If the spectrum contains excitations with a positive norm $\int\left[ |u_q|^2-|v_q|^2 \right]r\,\ud r$ but a negative eigenfrequency $\omega_q$, the stationary state is \emph{energetically unstable}. States that support modes with nonreal eigenfrequencies are referred to as \emph{dynamically unstable}, because the amplitude of a small perturbation associated with the excitation of a complex-frequency mode initially evolves exponentially in time [Eq.~(\ref{eq:oscillation})]. In the case of a multiquantum vortex, dynamical instability typically signifies that the vortex is unstable against splitting into singly quantized vortices~\cite{Mottonen2003,Shin2004,Huhtamaki2006b,Mateo2006,Isoshima2007,Takahashi2009,Kuopanportti2010b,Kuwamoto2010}. 

In the numerics, we measure length in units of the radial harmonic oscillator length $a_r=\sqrt{\hbar/\left(m\omega_r\right)}$ and energy in units of $\hbar\omega_r$ and normalize the dimensionless order parameter to unity. With these scalings, the dimensionless interaction strength becomes $\vv{g}=\sqrt{8\pi}N a/a_z$. In experiments, $\vv{g}$ lies typically between $10^2$ and $10^6$. 

\subsection{\label{subsc:plug_theory} Vortex stabilization with an optical plug}

As our first topic, we study the application of a repulsive plug potential as a means of dynamically stabilizing vortices with large winding numbers. In the case of giant vortices, dynamical instabilities correspond to deformations of the vortex core, and thus the quasiparticle amplitudes of dynamically unstable excitations are nonvanishing within the core region. Consequently, the instabilities of an axisymmetric giant vortex can be suppressed with a sufficiently strong and wide plug potential along the symmetry axis of the trap, since it increases the energy of such excitations. We assume that the plug has a Gaussian profile, \ie,
\begin{equation}\label{eq:plug}
V_\mathrm{plug}(r) = A e^{-r^2/d^2},
\end{equation}
where $A$ denotes the amplitude and $d$ is the beamwidth. The dynamical stabilization of a giant vortex was briefly investigated in Ref.~\cite{Kuopanportti2010b} in the case of an infinitely hard step-function potential, and the width of the step required to stabilize the vortex was found to be somewhat smaller than the size of the vortex core. 

In order to relate the parameters in Eq.~(\ref{eq:plug}) to experimental quantities, let us derive the expression from atomic properties. If the laser is operated at a frequency $\omega$ and the resulting electric field is denoted by $\vect{E}$, the potential experienced by an atom is given by $V_{\mathrm{plug}}=-\alpha(\omega)\overline{\vect{E}^2}/2$, where $\alpha(\omega)$ is the dynamical polarizability and $\overline{(\cdot)}$ denotes the time average. The spatial dependence in Eq.~(\ref{eq:plug}) follows from the spatial profile of the electric field, which for a focused laser beam is taken to be Gaussian. The polarizability is given by the Kramers--Heisenberg formula~\cite{Pethick2008}
\begin{equation}\label{eq:polarizability}
\alpha(\omega)=\frac{2}{\hbar}\sum_{e}\frac{\omega_{eg}|\matel{e}{\hat{\bm{\varepsilon}}\cdot\vect{d}}{g}|^2}{\omega_{eg}^2-\omega^2},
\end{equation}
where $\hat{\bm{\varepsilon}}$ is a unit vector in the direction of the electric field, $\vect{d}$ is the electric dipole-moment operator, and we label the ground state by $g$ and the excited states by $e$. The transition energies are expressed as $\hbar\omega_{eg}=\epsilon_e - \epsilon_g > 0$. We assume that the light is detuned far from the relevant atomic resonance at $\omega_0$ such that the detuning frequency $\omega-\omega_0$ is much larger than the natural decay rate of the corresponding excited state. For the two most common BEC species, ${}^{23}$Na and ${}^{87}$Rb, $\omega_0$ corresponds to the $n^2\mathrm{S}_{1/2}\rightarrow n^2\mathrm{P}_{3/2}$ transition, \ie, to the $\mathrm{D}_2$ line, with angular frequencies given respectively by $3.20\times 10^{15}\ \textrm{Hz}$ ($n=3$)~\cite{Juncar1981} and $2.41\times 10^{15}\ \textrm{Hz}$ ($n=5$)~\cite{Ye1996}. Thus, we limit the summation in Eq.~(\ref{eq:polarizability}) to the single hyperfine manifold of the $n^2\mathrm{P}_{3/2}$ level and neglect the hyperfine splitting between the different transitions, which yields
\begin{equation}\label{eq:polarizability2}
\alpha(\omega)=\frac{2}{\hbar}\frac{\omega_0}{\omega_0^2-\omega^2}\sum_{n^2\mathrm{P}_{3/2}} |\matel{e}{\hat{\bm{\varepsilon}}\cdot\vect{d}}{g}|^2\equiv\frac{2}{\hbar}\frac{\omega_0 }{\omega_0^2-\omega^2}|\vect{d}_\mathrm{eff}|^2,
\end{equation}
where we have labeled the remaining sum in terms of the effective dipole moment $\vect{d}_{\mathrm{eff}}$. The actual value of $|\vect{d}_{\mathrm{eff}}|$ depends on the atomic ground state and the polarization of the light but can, nevertheless, be readily evaluated~\cite{Steck2009a,Steck2009b}. In the case of ${}^{87}$Rb atoms in an $F=1$ hyperfine state and $\pi$-polarized light, one obtains $|\vect{d}_{\mathrm{eff}}|=2.44e_0 a_\mathrm{B}$, where $e_0$ is the electron charge and $a_\mathrm{B}$ is the Bohr radius. If we further assume that the laser operates at 660 nm (a typical experimental value) and relate the maximum electric field to the power $P$ of the laser, $|\vect{E}|^2_{\mathrm{max}} =2P/\left(\pi c \varepsilon_0 d^2\right)$, we find
\begin{equation}\label{eq:amplitude}
V_{\mathrm{plug}}(r)=  k_\mathrm{B} \times 73\,\mu \mathrm{K} \, \left( \frac{1\ \mu\mathrm{m}}{d} \right)^2\left( \frac{P}{1\ \mathrm{mW}} \right)e^{-\frac{r^2}{d^2}}.
\end{equation}

To study the dynamical stabilization of giant vortices numerically, we solve the stationary GP equation and the Bogoliubov equations for different values of the winding number $\kappa$, interaction strength $g$, plug amplitude $A$, and beamwidth $d$ and assess the dynamical stability of the corresponding stationary state. We use finite difference methods and solve the stationary GP equation using successive over-relaxation. The Bogoliubov equations are solved using the {\footnotesize LAPACK} numerical library in {\footnotesize MATLAB}~\cite{matlab}.

\subsection{\label{subsc:pump_theory}Optical plug in vortex pumping}

As our second topic, we demonstrate that by utilizing a sufficiently strong plug potential, the vortex pump method~\cite{Mottonen2007} can be used to create an unsplit giant vortex with a very large winding number. To this end, we study the temporal evolution of a BEC during vortex pumping in a case where the harmonic trap is combined with a strong Gaussian potential of the form of Eq.~(\ref{eq:plug}). Since the vortex pump makes explicit use of the spin degree of freedom of the BEC, we no longer assume the atomic spins to be fixed. Instead, we consider a spinor BEC with a hyperfine spin $F=1$ and model it with the time-dependent spin-1 GP equation~\cite{Ohmi1998,Ho1998}
\begin{eqnarray}
\label{eq:spinorGPE}
i\hbar\partial_t\spinor(r,\phi,t) &=& \Big[ {\mathcal H} + \mu_\mathrm{B} g_F \vect{B}(t)\cdot \fmatrix + c_0\spinor^\dagger\spinor \nonumber \\
&&  + c_2  \big( \spinor^\dagger \fmatrix \spinor \big) \cdot \fmatrix \Big] \spinor(r,\phi,t),
\end{eqnarray}
where $\fmatrix=({\mathcal{F}}_x,{\mathcal{F}}_y,{\mathcal{F}}_z)^\mathrm{T}$ consists of the generators of the spin rotation group $\mathrm{SO}(3)$ and $\spinor=\left(\wf_1,\wf_0,\wf_{-1}\right)^\mathrm{T}$ is the three-component spinor order parameter written in the eigenbasis of ${\mathcal{F}}_z$. Moreover, $\mu_\mathrm{B}$ is the Bohr magneton, $g_F$ is the Land\'{e} $g$~factor, and $c_0 = \sqrt{8\pi}\hbar^2\left(a_{0}+2a_{2}\right)/\left(3ma_z\right)$ and $c_2 = \sqrt{8\pi}\hbar^2(a_{2}-a_{0})/\left(3ma_z\right)$ are the coupling constants related to $s$-wave scattering lengths $a_{0}$ and $a_{2}$ for different spin channels.
As in Eq.~(\ref{eq:tGPE}), we have again assumed that the BEC is pancake-shaped and integrated out the $z$ dependence of the order parameter. 

\begin{figure}
\begin{center}
\includegraphics[
  width=230pt,
  keepaspectratio]{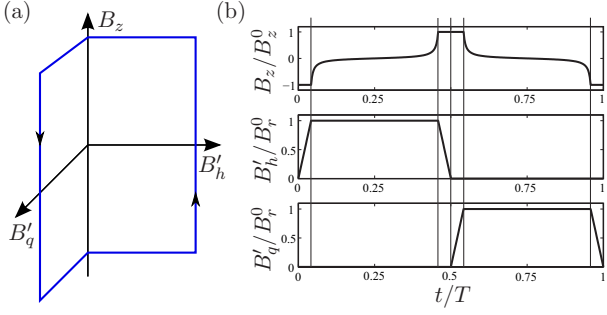}
\end{center}
\caption{\label{fig:cycle}
(Color online) (a) Control cycle of the vortex pump simulation in the magnetic-field parameter space.  Here, $B_q'= | \vect{B}_q|/r$ and $B_h'=|\vect{B}_h|/r$ denote the radial derivatives of the quadrupole and hexapole field magnitudes. (b) Temporal changes of the control parameters during the pumping cycle given in Eq.~(\ref{eq:magnetic_field}). The common maximum value of $B_q'$ and $B_h'$ is denoted by $B_r^0$.}
\end{figure}

In the vortex pump, the spin degree of freedom of the condensate is controlled locally by slowly tuning the external magnetic field $\vect{B}(t)$ in a cyclic manner such that the system acquires a fixed amount of vorticity in each control cycle. The efficient operation of the pump requires that sudden spin flips due to Landau--Zener transitions are insignificant, and therefore, the control cycle should be sufficiently adiabatic~\cite{remark0}. In our simulation, we use a cycle which is identical to the one employed in Ref.~\cite{Mottonen2007}, consisting of a homogeneous bias field in the $z$ direction and alternating quadrupole and hexapole fields in the $xy$ plane. Denoting the quadrupole and hexapole fields by $\vect{B}_\mathrm{q}^0 = B_r^0 r \left[ \cos(\phi)\hat{\vect{x}} - \sin(\phi)\hat{\vect{y}}\right]$ and $\vect{B}_\mathrm{h}^0 = B_r^0 r\big[ \cos(2\phi)\hat{\vect{x}}-\sin(2\phi)\hat{\vect{y}}\big]$, the exact form of the control cycle becomes
\begin{equation}\label{eq:magnetic_field}
\vect{B}= \left\{ \begin{array}{ll} -B_z^0\hat{\vect{z}} + \frac{t}{T_1}\vect{B}_\mathrm{h}^0, & 0 \leq t < T_1, \\
f\left(\frac{t-T_1}{T_2}\right)B_z^0 \hat{\vect{z}}+\vect{B}_\mathrm{h}^0, & T_1 \leq t < T_1+T_2, \\
B_z^0\hat{\vect{z}} + \frac{2t-T}{2T_1}\vect{B}_\mathrm{h}^0,& T_1 + T_2 \leq t < \frac{T}{2}, \\
B_z^0\hat{\vect{z}} + \frac{2t-T}{2T_1}\vect{B}_\mathrm{q}^0, &\frac{T}{2} \leq t < \frac{T}{2}+T_1, \\
-f\left(\frac{t-3T_1-T_2}{T_2}\right)B_z^0 \hat{\vect{z}}+\vect{B}_\mathrm{q}^0, &\frac{T}{2} + T_1 \leq t < T - T_1, \\
-B_z^0\hat{\vect{z}} + \frac{T-t}{T_1}\vect{B}_\mathrm{q}^0,& T-T_1 \leq t \leq T, \end{array} \right.
\end{equation}
where the function 
\begin{equation}\label{eq:flip_function}
f(x)=\frac{5 a_rB_{r}^0}{B_z^0} \tan\left[\left(2x-1\right)\arctan\left(\frac{B_z^0}{5a_r B_{r}^0}\right)\right],
\end{equation}
$0\leq x\leq 1$, is such that the spin at a distance of $5 a_r$ from the $z$ axis is reversed with a constant speed. 
Here, $T_1$ is the ramping time of the multipole fields, $T_2$ is the bias-field inversion time, and $T=4T_1+2T_2$ is the total period of the cycle. The control cycle is visualized in Fig.~\ref{fig:cycle}. In order to make the effect of the plug potential transparent, we set the system parameters identical to those used in the original simulation of Ref.~\cite{Mottonen2007} except for the plug potential, which we choose to be significantly stronger. Thus, we aim at achieving a significantly higher winding number than the value $\kappa=8$ reported in Ref.~\cite{Mottonen2007}. Before beginning the pumping, we use the method of successive over-relaxation to find the ground state of the condensate with the magnetic field set to its initial configuration, $\vect{B}(t=0)=-B_z^0\hat{\vect{z}}$. Equation~(\ref{eq:spinorGPE}) is then numerically integrated using the Strang splitting scheme.

\section{\label{sc:results}Results}

\subsection{\label{subsc:plug_results} Vortex stabilization with an optical plug}

By solving the stationary GP and Bogoliubov equations numerically, we have determined the values of the plug amplitude $A$ and width $d$ that are sufficient to render the multiquantum vortex states dynamically stable at different winding numbers $\kappa$ and dimensionless interaction strengths $\vv{g}=\sqrt{8\pi}N a/a_z$. The value of $\vv{g}$ has been restricted to the range $0\leq\vv{g}\leq 2000$, but this should not limit the generality of our results.

\begin{figure}
\begin{center}
\includegraphics[
  width=210pt,
  keepaspectratio]{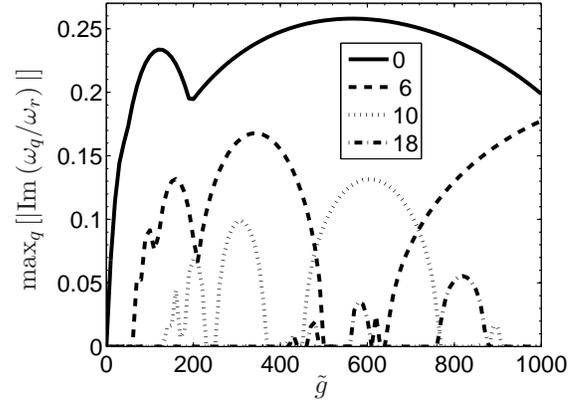}
\end{center}
\caption{\label{fig:maxims} Maximum imaginary part of the eigenfrequencies for a 10-quantum vortex as a function of the interaction strength $\vv{g}$ for different values of the plug amplitude $A$ given in the inset. The width of the plug is set to $d=3a_r$. The imaginary parts vanish completely for $A\geq 26.0\hbar\omega_r$. The dimensionless interaction strength is given by $\vv{g}=\sqrt{8\pi}N a/a_z$, where $N$ is the particle number, $a$ the $s$-wave scattering length, and $a_z=\sqrt{\hbar/\left(m\omega_z\right)}$ the axial harmonic oscillator length.
}
\end{figure}

\begin{figure}
\begin{center}
\includegraphics[
  width=220pt,
  keepaspectratio]{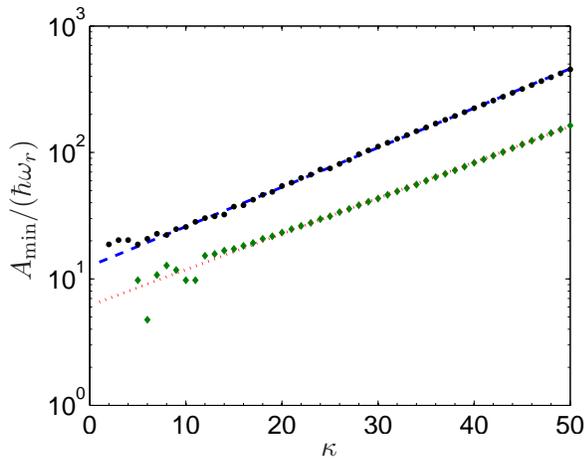}
\end{center}
\caption{\label{fig:plug-fixed_width} (Color online) Minimum amplitude $A_\mathrm{min}$ [see Eq.~(\ref{eq:amplitude})] required to stabilize vortices with a given winding number $\kappa$. The black dots indicate values that stabilize the vortices in the whole interval $0\leq\vv{g}\leq 2000$, and the dashed blue line is the least-squares fit $A_\mathrm{min}=13 \hbar \omega_r \exp(0.072\kappa)$. The green diamonds show the values that stabilize the vortex for fixed $\vv{g}=250$, and the dotted red line represents the fit $A_\mathrm{min}=6.2 \hbar \omega_r \exp(0.065\kappa)$. In both cases, the width of the plug is set to $d=3a_r$.}

\end{figure}

\begin{figure}
\begin{center}
\includegraphics[
  width=220pt,
  keepaspectratio]{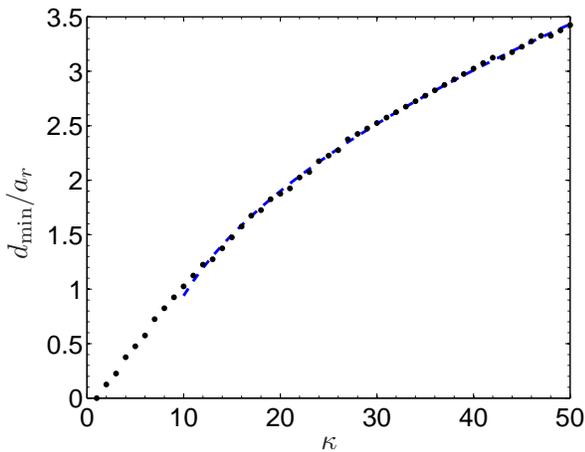}
\end{center}
\caption{\label{fig:plug-fixed_amp} (Color online) Minimum beamwidth $d_\mathrm{min}$ [see Eq.~(\ref{eq:amplitude})] required to stabilize vortices with a given winding number $\kappa$ in the whole interval $0\leq\vv{g}\leq 2000$. The dotted curve is a two-parameter fit to a square-root function for $\kappa \geq 10$, given by $d_\mathrm{min}/a_r = 0.52 \sqrt{\kappa-6.8}$. The amplitude of the plug is set to $A=200\hbar\omega_r$.}

\end{figure}

Figure~\ref{fig:maxims} shows the strongest dynamical instability of a 10-quantum vortex as a function of $\vv{g}$, $0 \leq \vv{g} \leq 1000$, for different values of the amplitude $A$. In the absence of the plug, the vortex is dynamically unstable in the whole interval $0 < \vv{g} \leq 1000$, but as the strength of the plug increases, regions of dynamical stability begin to appear, until eventually the unstable modes vanish completely.  

In Fig.~\ref{fig:plug-fixed_width}, we plot, as a function of $\kappa$, the limiting plug amplitude $A_\mathrm{min}$ above which the $\kappa$-quantum vortex is dynamically stable in the whole interval $0 \leq \vv{g} \leq 2000$. Here, the beamwidth is fixed at $d=3a_r$, but quantitatively similar behavior is found for other values of $d$ as well. The stabilizing amplitude $A_\mathrm{min}$ increases exponentially with $\kappa$ for sufficiently large winding numbers, and ordinary least-squares fitting yields the expression $A_\mathrm{min}=13 \hbar\omega_r \exp(0.072 \kappa)$. Roughly speaking, the exponential dependence follows from requiring that the plug potential exceeds a certain limiting strength $V_0$ inside the whole core region $r\leq r_\mathrm{c}$, where $r_\mathrm{c}$ denotes the vortex core radius. The plug is weakest at $r=r_\mathrm{c}$, and thus the limiting condition becomes $A_\mathrm{min}\exp(-r_\mathrm{c}^2/d^2)=V_0$. It was found in Ref.~\cite{Kuopanportti2010a} that for a sufficiently large $\kappa$, $r_\mathrm{c}/a_r \propto \sqrt{\kappa}$. Neglecting the dependence of $V_0$ on $\kappa$, we get $\log A_\mathrm{min} \propto \kappa$~\cite{remark1}. The irregular behavior at $\kappa< 10$ is explained by noting that the dependence of $r_\mathrm{c}$ on $\kappa$ deviates significantly from the square-root form if $\vv{g} \gg \kappa^2$~\cite{Kuopanportti2010a}. 

The obtained values of $A_\mathrm{min}$ are small enough to be realizable with commercial lasers: If we use the values $a = 4.7\ \textrm{nm}$ and $m=1.44 \times 10^{-25}\ \textrm{kg}$ corresponding to ${}^{87}$Rb atoms~\cite{Pethick2008}, choose $\left(\omega_r,\omega_z\right) = 2\pi \times \left(8,90\right)\,\textrm{Hz}$~\cite{Neely2010}, and use the plug potential in Eq.~(\ref{eq:amplitude}), we find that $d=3a_r\approx 11\ \mu\mathrm{m}$ and that $A_\mathrm{min}=454 \hbar\omega_r$ at $\kappa=50$ is obtained with the laser power $P=0.31\ \textrm{mW}$. Values of this order have been used in various experiments~\cite{Abo-Shaeer2001,Raman2001,Simula2005,Neely2010}.

The smallest beamwidth $d_\mathrm{min}$ required to stabilize the multiquantum vortices with $0\leq \vv{g}\leq 2000$ and a given winding number $\kappa$ is displayed in Fig.~\ref{fig:plug-fixed_amp} as a function of $\kappa$. The amplitude is fixed at $A=200 \hbar \omega_r$. The minimum width $d_\mathrm{min}$ increases as a square-root function of $\kappa$. This can again be understood by assuming that the plug is stabilizing when its strength exceeds the threshold value $V_0$ for $r \leq r_\mathrm{c}$. Thus, we require $A \exp(-r_\mathrm{c}^2/d_\mathrm{min}^2)=V_0$, which yields $d_\mathrm{min} \propto r_\mathrm{c} \propto \sqrt{\kappa}$.

\subsection{\label{subsc:pump_results}Optical plug in vortex pumping}

To demonstrate the efficiency of the stabilizing plug, we have computed the time evolution of the spinor order parameter $\spinor$ from Eq.~(\ref{eq:spinorGPE}) during vortex pumping. The optical plug has been chosen to have the amplitude $A=100\hbar\omega_r$ and width $d=3a_r$, and thus it is considerably stronger than that used in Ref.~\cite{Mottonen2007} ($A=10\hbar\omega_r$ and $d=2a_r$). Otherwise, the system is identical to the one considered in Ref.~\cite{Mottonen2007}, where unsplit vortices up to the winding number $\kappa=8$ were reached. More specifically, the dimensionless interaction parameters are $\vv{c}_0 = m N c_0 / \hbar^2 = 250$ and $\vv{c}_2= m N c_2 / \hbar^2 = -0.01 \vv{c}_0$ (which corresponds to ${}^{87}$Rb), the magnetic field strengths read $B_z^0  = 40 \hbar \omega_r /\left(\mu_\mathrm{B}|g_F|\right)$ and $B_r^0 =  \hbar \omega_r  / \left(a_r \mu_\mathrm{B}|g_F|\right)$, and the multipole-field-ramping and bias-field-inversion times are given by $T_1 = 10/\omega_r$ and $T_2=160/\omega_r$ such that the cycle period is $T=360/\omega_r$ [see Eq.~(\ref{eq:magnetic_field})]. The Land\'{e} $g$ factor $g_F$ is taken to be negative. Assuming again that the BEC consists of ${}^{87}$Rb atoms and $\left(\omega_r,\omega_z\right) = 2\pi \times \left(8,90\right)\,\textrm{Hz}$~\cite{Neely2010}, the simulation parameters correspond to $a_r \approx 3.8 \ \mu\mathrm{m}$, $N\approx 10^4$, $B_z^0 \approx 46\textrm{ nT}$, and $B_r^0 \approx 0.3\textrm{ nT}/\mu\mathrm{m}$.

\begin{figure}
\begin{center}
\includegraphics[
  width=220pt,
  keepaspectratio]{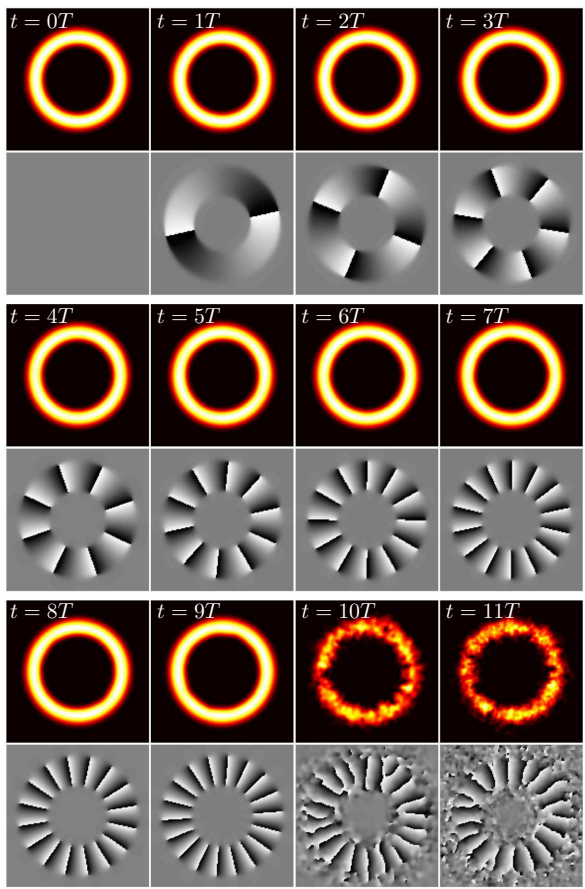}
\end{center}
\caption{\label{fig:pump1} (Color online) Areal particle density (upper panels) and complex phase (lower panels) of the component $\wf_{-1}$ of the spinor order parameter in the $xy$ plane at integer multiples of the vortex pumping period $T=360/\omega_r$, with a plug potential of width $d=3a_r$ and amplitude $A=100\hbar\omega_r$. The field of view in each panel is $18\,a_r\times 18\,a_r$. 
}
\end{figure}

\begin{figure}
\begin{center}
\includegraphics[
  width=220pt,
  keepaspectratio]{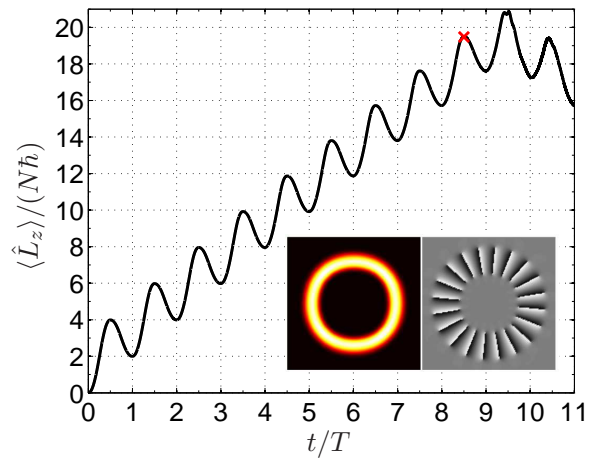}
\end{center}
\caption{\label{fig:pump2} (Color online) Axial angular momentum of the condensate as a function of time during the vortex pump simulation presented in Fig.~\ref{fig:pump1}. The inset shows the particle density and phase of the 20-quantum vortex (field of view $18\,a_r\times 18\,a_r$) at the time instant $t=8.5T$ marked with the red cross.
}
\end{figure}

In Fig.~\ref{fig:pump1}, we present the squared modulus and complex phase of the order parameter component $\wf_{-1}$ at integer multiples of the pumping period $T$. The accumulation of two quanta of vorticity per cycle is clearly visible in the phase field. Figure~\ref{fig:pump2} shows the corresponding time-dependence of the total axial angular momentum $\langle{\hat{L}_z}\rangle$ of the BEC. The notable deviation of $\langle{\hat{L}_z}\rangle$ from the ideal value $2k N \hbar$ after $k \approx 8$ pumping cycles is due to excitations away from the instantaneous eigenstate the pump is operated in. These excitations are imperceptible in Fig.~\ref{fig:pump1} during the nine first cycles, so that a clear 18-quantum vortex state is observed at $t=9T$. In fact, a symmetric 20-quantum vortex is obtained in the middle of the ninth cycle at $t=8.5T$. We have also checked that the plug amplitude can be ramped down without destroying the giant vortex.

After $t=9T$, the dissociation of the giant vortex begins in spite of the plug, causing single vortices to move out of the condensate. This is also manifested in Fig.~\ref{fig:pump2} by the decrease of the axial angular momentum. Moreover, by using initial states with higher vorticity, we have verified that the splitting after $t=9T$ is not caused by the accumulation of numerical errors in the temporal evolution and that a higher winding number is reached by further increasing the strength of the optical plug.  We have also observed that the abrupt removal of the plug and the addition of weak random noise to the state at $t=8.5T$ or $t=9T$ eventually results in the splitting of the 18- or 20-quantum vortex with a fourfold-symmetric splitting pattern introduced in Ref.~\cite{Kuopanportti2010b}.

\section{\label{sc:discussion}Discussion}

In this article, we have studied the dynamical stabilization of giant vortices in harmonically trapped BECs by applying a Gaussian-shaped repulsive plug potential along the symmetry axis of the trap. We found that vortices with large winding numbers can be stabilized with plug profiles that should be routinely achievable with commercial lasers. Although the detailed behavior of the dynamical instabilities as a function of the plug parameters turned out to be complicated (Fig.~\ref{fig:maxims}), the overall criterion for the stabilization could be explained in simple terms of filling the vortex core with a sufficiently high potential barrier. 

We also performed a simulation of the vortex pump which indicated that a giant vortex can be created by pumping if a sufficiently strong optical plug is utilized and the temperature is kept low enough such that dissipation effects due to the thermal gas are negligible.  In the simulation, the duration of the control cycle was kept constant throughout the creation process. In fact, it is possible to gradually increase the pumping speed as vorticity accumulates into the BEC and still retain the adiabaticity of the process~\cite{Kuopanportti2010a}. By making use of this possibility, one could increase the winding number beyond the value $\kappa = 20$ reached in our simulation before the vortex splits. Furthermore, one could also employ a different pumping scheme~\cite{Xu2008,Xu2010} which increases the vorticity of the $F=1$ condensate by $4 h/m$ per cycle instead of $2 h/m$ associated with the control cycle used in our calculation.

It should be pointed out that according to our analysis in Sec.~\ref{subsc:plug_results}, the optical plug employed in the vortex pump simulation should render all $\kappa$-quantum vortices dynamically stable at $\tilde{g}=250$ up to $\kappa=42$, but here the splitting was observed already before reaching $\kappa = 22$. The discrepancy is likely due to the fact that the stability analysis of Sec.~\ref{subsc:plug_results} concerns a perfectly spin-polarized BEC in a pure harmonic trap. Thus, it can be strictly applied to the spinor BEC in vortex pumping only when the external magnetic field consists of the strong homogeneous bias field, \ie, at the start and middle point of each pumping cycle. The additional instabilities related to the spinor nature of the BEC and the presence of the multipole magnetic field have not been taken into account, and hence the stability analysis of Sec.~\ref{subsc:plug_results} only provides lower limits of the stabilizing plug parameters for the vortex pump~\cite{remark2}. Nonetheless, the qualitative behavior $A_\mathrm{min} \propto\exp(\alpha\kappa)$ for some constant $\alpha$ is still anticipated in the spinor BEC. 

\begin{acknowledgments}
The authors acknowledge the Academy of Finland, the Emil Aaltonen Foundation, the V\"ais\"al\"a Foundation, and Finnish Academy of Science and Letters for financial support. V.~Pietil\"a and T.~P.~Simula are appreciated for insightful discussions. V.~Pietil\"a is also acknowledged for developing the original computer program used for obtaining the results of Sec.~\ref{subsc:pump_results}.
\end{acknowledgments}

\bibliography{plug-revtex}

\begin{thebibliography}{53}
\bibitem[{com()}]{remark0}
\bibinfo{note}{Note that neither Eq.~(\ref{eq:spinorGPE}) nor our simulation explicitly assumes adiabaticity.}
\bibitem[{com()}]{remark2}
\bibinfo{note}{The bias and multipole magnetic fields together yield an approximately harmonic potential which acts to reduce the effective trapping frequency $\omega_r$ and beamwidth $d$. With our parameter values, however, the decrease in $d$ is only 0.6\%, and thus this rescaling alone cannot explain the splitting at a smaller value of $\kappa$ than expected.}
\end{thebibliography}
\end{document}